\documentclass[sigconf,nonacm]{acmart} 

\renewcommand\footnotetextcopyrightpermission[1]{}

\acmConference[SIGCSE TS '26]{ACM Technical Symposium on Computer Science Education}{March 18–21, 2026}{Seattle, WA, USA}
\acmDOI{}
\acmISBN{}
\acmPrice{}

\usepackage{draftwatermark}
\SetWatermarkText{\textcolor{blue}{Preprint}}
\SetWatermarkScale{0.2}
\SetWatermarkColor[gray]{0.85}
\SetWatermarkAngle{45}

\AtBeginDocument{%
  \fancyhead[L]{\textit{\textcolor{blue}{Preprint — Accepted at ACM SIGCSE TS 2026}}}%
  \fancyhead[R]{\thepage}%
}
\AtBeginDocument{%
  }

\usepackage{subcaption}

\usepackage{tabularx}
\usepackage{booktabs}
\usepackage{geometry}

\begin{document}

\title{Developing a Decolonial Mindset for Indigenising Computing Education (CE)}


\author{Jianhua Li}
\email{jack.li@deakin.edu.au}
\affiliation{%
  \institution{Deakin University}
  \city{Melbourne}
  \state{Vic}
  \country{Australia}
}
\author{Yin Paradies}
\email{yin.paradies@deakin.edu.au}
\affiliation{%
  \institution{Deakin University}
  \city{Melbourne}
  \state{Vic}
  \country{Australia}
}
\author{Trina Myers}
\email{trina.myers@deakin.edu.au}
\affiliation{%
  \institution{Deakin University}
  \city{Melbourne}
  \state{Vic}
  \country{Australia}
}
\author{Robin Doss}
\email{robin.doss@deakin.edu.au}
\affiliation{%
  \institution{Deakin University}
  \city{Melbourne}
  \state{Vic}
  \country{Australia}
}
\author{Armita Zarnegar}
\email{azarnegar@swin.edu.au}
\affiliation{%
  \institution{Swinburne University of Technology}
  \city{Melbourne}
  \state{Vic}
  \country{Australia}
}
\author{Jack Reis}
\email{jack@baidam.com.au}
\affiliation{%
  \institution{Baidam Solutions}
  \city{Brisbane}
  \state{Queensland}
  \country{Australia}
}

\ccsdesc[500]{Applied computing~Collaborative learning}
\keywords{Curriculum Transformation, Underrepresentation, Co-Development}

\begin{abstract}
The persistent underrepresentation of First Peoples in computing education reflects enduring legacies of colonialism embedded in curricula, pedagogies, and digital infrastructures. This position paper introduces the \textbf{Decolonial Mindset Stack (DMS)}, a seven-layer conceptual framework that scaffolds computing educators through stages of mindset transformation: \textbf{Recognition, Reflection, Reframing, Reembedding, Reciprocity, Reclamation}, and \textbf{Resurgence}. Each layer is grounded in Freirean critical pedagogy and Indigenous methodologies, aligning with relational lenses such as “About Me,” “Between Us,” and “By Us.” Together, these dimensions foster self-reflexivity, relational accountability, and Indigenous sovereignty in computing education (CE). The DMS reframes underrepresentation as an outcome of systemic exclusion and a call to action for reimagining how CE can be co-developed with and led by First Peoples communities. We present the theoretical grounding, illustrate implementation pathways, and argue that indigenisation is not an endpoint but a sustained ethical commitment to transformative justice in CE.
\end{abstract}

\maketitle
\section{Introduction}

The persistent underrepresentation of First Peoples in computing education (CE) is a pressing global issue that underscores the need for systemic change \cite{peters2024sustainability}. In Australia, First Nations students make up only 1.8\% of higher education students, with even fewer participating in STEM fields \cite{aihw2023-ub}. Similarly, in the United States, Native students receive just 1.4\% of associate degrees and 0.6\% of Bachelor’s degrees in computer science \cite{blake2023indigeneous}. In Canada, Indigenous youth face significant barriers to postsecondary STEM participation, with lower enrolment and completion rates compared to non-Indigenous peers \cite{bruce2012literature}. Consequently, this underrepresentation limits access to technological opportunities, perpetuates socio-economic disparities, and restricts the diversity of perspectives critical for innovative computing solutions \cite{chari2024bridging}. The 2016 ABS data revealed only 231 Indigenous men and 84 women were employed as `IT support technicians,' underscoring the profound underrepresentation in Australia \cite{braue2021too}. These disparities reflect broader patterns of exclusion rooted in colonial legacies.

Colonial influences have profoundly fashioned CE, embedding Western-centric content, pedagogies, and infrastructures that often marginalise Indigenous knowledge and ways of learning \cite{geyser2025decoloniality}. Curricula typically emphasise Western computational paradigms, such as algorithmic thinking, while sidelining Indigenous epistemologies that value relationality and contextual understanding. Pedagogical approaches, predominantly online lecture-based, may not resonate with Indigenous learning styles, which often prioritise oral traditions and community engagement. Infrastructural barriers, including limited access to technology in many Indigenous communities, further hinder participation and success in CE \cite{prayaga2017digital}.

To confront this underrepresentation, there is a critical need for educators to adopt relational, reflective, and sovereignty-oriented mindsets. This involves recognising and challenging colonial biases, building meaningful relationships with Indigenous communities, and supporting Indigenous sovereignty in educational spaces \cite{khalifa2019toward}. Such transformation is essential for creating CE that is inclusive, culturally sustaining, and empowering for Indigenous students. Hence, we introduce the \textbf{Decolonial Mindset Stack (DMS)}, a conceptual framework designed to guide educators through the process of decolonising and indigenising CE. The DMS consists of seven layers: \textbf{Recognition, Reflection, Reframing, Reembedding, Reciprocity, Reclamation, and Resurgence}, each paired with relational lenses that emphasise different aspects of Indigenous engagement, from ``About Me'' to ``By Us.'' Grounded in Freirean critical pedagogy and Indigenous methdologies \cite{freire2020pedagogy}, the DMS provides a structured pathway to develop the mindsets and practices necessary for transformative change. 

Educators can use the DMS as a tool to navigate the complexities of decolonising CE. The remainder of this paper is structured as follows. Section~\ref{sec:context} outlines the contextual background, followed by the theoretical foundations in Section~\ref{sec:theory}. Section~\ref{sec:dms} presents the DMS in detail, explaining each layer and relational lens in terms of its role in fostering a decolonial mindset. Section~\ref{sec:discussion} discusses the framework’s significance, scope, intended audience, ethical considerations, and implications for equity and reflective practice. Finally, Section~\ref{sec:conclusion} concludes the paper  and presents a call to action, advocating decolonisation as a pathway to tackling the persistent underrepresentation in computing and technology education.

\section{Background and Motivation}
\label{sec:context}
Recent scholarship highlights the potential of narrative approaches to reframe computing as a decolonising force, fostering equitable and sustainable educational practices \cite{karetai2023decolonising, ali2020decolonising, carroll2023decolonizing, geyser2025decoloniality}. These works emphasise the integration of Indigenous perspectives, although they do not focus specifically on the underrepresentation of First Peoples. Carruthers examined publications on the indigenisation and decolonisation of CE between 2010 and 2023, concluding that the shared emphasis was on achieving cultural relevance \cite{carruthers2024indigenization}. Eglash \textit{et al.} argued that culture-based approaches to CE held promise for increasing the interest and engagement of underrepresented students \cite{eglash2013toward}. Maniapoto \textit{et al.} reported that First Peoples remain significantly underrepresented in CE in Aotearoa New Zealand \cite{maniapoto2025incorporating}, while Lachney \textit{et al.} contended that this underrepresentation is unlikely to improve without greater attention to culturally responsive computing \cite{lachney2021teaching}. 

The underrepresentation of First Peoples in CE results in a significant loss of diversity, perpetuating cycles of disadvantage by restricting access to high-paying technology careers and broader economic participation. This exclusion reinforces historical and systemic inequalities, diminishes cultural representation in technological development, and limits the diversity of thought critical for innovation in computing \cite{maniapoto2025incorporating,li2023adapting}. Alarmingly, persistent underrepresentation has even led to harmful technical failures, such as machine learning systems misclassifying Black Americans as gorillas due to insufficient training data \cite{gebru2020race}. Our work addresses this gap by transforming educational practices and mindsets to foster equitable inclusion.

\begin{figure}[ht!]
    \centering
    \includegraphics[width=0.48\textwidth]{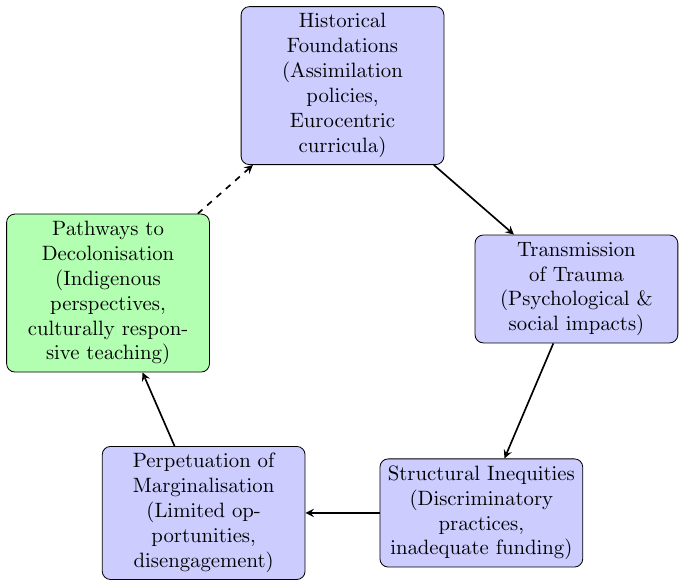}
    \caption{The systemic cycle of marginalisation in education begins with historical foundations and continues through structural and psychological impacts. However, the ``Pathways to Decolonisation” acts as a critical intervention point, disrupting this cycle by embedding Indigenous perspectives and culturally responsive teaching.}
     \Description{A diagram showing the systemic cycle of marginalisation in education, with an intervention point labeled 'Pathways to Decolonisation' that disrupts the cycle.}
    \label{fig:decolonial-cycle}
\end{figure}

\textbf{Coloniality} in education refers to the enduring legacy of colonial power structures that privilege Western knowledge systems while suppressing Indigenous epistemologies \cite{omodan2024roles}. The intergenerational cycle of colonial marginalisation in education (Fig.~\ref{fig:decolonial-cycle}) illustrates how colonial policies (e.g., forced assimilation and the disruption of traditional learning systems) severed cultural transmission and inflicted profound trauma on Indigenous peoples. This trauma continues across generations, diminishing cultural pride, self-esteem, and academic engagement. Critically, contemporary systems perpetuate these colonial legacies through underfunding, culturally irrelevant curricula, and discriminatory practices that limit educational opportunities. The combined effects of historical trauma and ongoing structural inequity create a self-reinforcing cycle, restricting educational achievement and social mobility while entrenching marginalisation. Breaking this cycle requires systemic transformation: integrating Indigenous knowledge systems, adopting culturally responsive pedagogies, and empowering communities through participatory governance to restore cultural identity.

While diversity initiatives, like scholarships or outreach programs, aim to increase Indigenous participation, they frequently fail to deal with the root causes of exclusion. Many of these efforts centre on surface-level solutions, such as boosting enrolment, without challenging the colonial structures embedded in educational systems \cite{gray2012increasing}. For instance, First Nations students who enter computing programs often encounter curricula and pedagogies that disregard their cultural values and knowledge systems, contributing to disproportionately high attrition rates. Data from \textbf{Universities Australia} shows that while Indigenous student enrolments in higher education grew from 9,490 in 2006 to 22,897 in 2020 (annual increase of 7.6\%), their attrition rate remained around 20\% during the same period \cite{jackson2022indigenous}. A national study further highlights this disparity, showing that only 47\% of Aboriginal and Torres Strait Islander students completed their degrees, compared to 74\% of non-Indigenous students \cite{hearn2021investigation}. These gaps persist because many diversity programs neglect the ongoing impacts of colonialism on Indigenous communities’ access to and experiences in education \cite{puritty2017without}. When Indigenous perspectives are excluded from the design and implementation of educational initiatives, the resulting programs often fail to align with their needs and priorities. This underscores the necessity of systemic, culturally responsive interventions to foster genuinely inclusive learning environments \cite{Li2025unlocking}.

To tackle these challenges, a strategic intervention is required: a transformation of educators’ mindsets toward decolonisation and indigenisation. This involves cultivating relational, reflective, and sovereignty-oriented mindsets that recognise colonial biases, engage meaningfully with Indigenous communities, and support their self-determination \cite{hunt2013engaging}. Such a transformation reorients CE to pivot Indigenous epistemologies, fostering environments where First Nations students can thrive. To this end, DMS is complementary to Indigenous scholarships and outreach programs, extending systemic equity and Indigenous empowerment.

\section{Theoretical Foundations}
\label{sec:theory}
The DMS for indigenising CE is grounded in Freirean Critical Pedagogy and Indigenous methodologies \cite{freire2020pedagogy}. These perspectives provide a robust foundation for addressing colonial legacies and fostering inclusive, culturally sustaining educational practices.

\subsection{Freirean Critical Pedagogy}
Paulo Freire's Critical Pedagogy, as articulated in \textbf{Pedagogy of the Oppressed} \cite{freire2020pedagogy}, emphasises \textbf{conscientisation} and \textbf{praxis} as tools for educational transformation. Conscientisation involves developing critical awareness of social and political realities, enabling educators and students to recognise and challenge oppressive structures, such as the exclusion of Indigenous perspectives in CE. This aligns with the DMS layers of Recognition, where educators acknowledge colonial influences, and Reflection, where they critically examine their own biases and complicity. Praxis, the iterative cycle of action and reflection, underpins the entire DMS framework, encouraging educators to actively transform curricula and pedagogies through ongoing engagement. Freire's critique of the ``banking model'' of education, where students are passive recipients, supports the DMS's call for dialogical, participatory learning that empowers Indigenous students.

\subsection{Indigenous Methodologies}
Indigenous methodologies hinge Indigenous ways of knowing, being, and doing, offering culturally relevant approaches to education. Key concepts include:

\begin{itemize}
    \item \textbf{Relationality}: Emphasises the foundational principle that all elements of existence (people, land, ancestors, spirits, plants, animals) are interconnected and mutually responsible \cite{mbah2024considerations}. Relational learning is a process where knowledge is acquired through dynamic, reciprocal relationships anchored in connections to community, land, ancestors, and cosmos. 
    \item \textbf{Sovereignty}: Advocates for Indigenous control over their knowledge and data, aligning with the Resurgence layer, where communities lead their educational futures \cite{oguamanam2019indigenous}. Knowledge Sovereignty is the right of Indigenous peoples to control, protect, and ethically steward their ancestral knowledge systems, free from exploitation, appropriation, or external authority.
    \item \textbf{Two-Eyed Seeing}: is a Mi'kmaw guiding principle for integrating Indigenous and Western knowledge systems in learning, using the strengths of both ``eyes'' to gain deeper wisdom without hierarchy or assimilation. Mi'kmaw Elder Albert Marshall says, ``\emph{To see with one eye Indigenous wisdom, and with the other Western science—using both together for the benefit of all.}'' This approach integrates Indigenous and Western knowledge for a holistic perspective, guiding the DMS's balanced approach to indigenisation \cite{bartlett2012two}.
    \item \textbf{8 Ways of Learning}: is an Aboriginal Australian pedagogical framework that integrates Indigenous knowledge systems into education through eight interconnected, holistic practices. Developed by Prof. Tyson Yunkaporta and the NSW Department of Education, it centres culture, community, and Country as foundations for teaching and learning, informing the Reembedding layer to make computing curricula culturally relevant \cite{yunkaporta2009reclaiming}.
\end{itemize}

Freirean Critical Pedagogy and Indigenous methodologies shape the DMS by providing critical and cultural lenses for decolonisation. Conscientisation drives the Recognition layer, raising awareness of Indigenous histories and exclusion, while praxis informs the iterative process across all layers, particularly Reframing and Reembedding, where colonial structures are challenged and Indigenous content is integrated. Indigenous methodologies direct the relational and sovereignty-oriented layers: relationality underpins Reciprocity, ensuring community partnerships; sovereignty forms Resurgence, empowering Indigenous leadership; and the 8 Ways of learning enriches Reembedding, implanting culturally resonant pedagogies. Two-Eyed Seeing supports a balanced integration of knowledge systems, ensuring the DMS is both inclusive and practical for CE.

\section{The DMS Framework}
\label{sec:dms}
\subsection{Development of the DMS}
The DMS serves as a layered scaffold for educators seeking to decolonise computing education (CE). It outlines seven mindset shifts that guide educators from initial awareness of Indigenous histories and exclusion to supporting Indigenous sovereignty in digital spaces. In brief, Layers 1 to 3 focus on decolonisation, creating opportunities for indigenisation at Layers 4 to 7. Each layer is paired with a relational lens, ranging from About Me'' to By Us'', reflecting the evolving nature of the educator's engagement with Indigenous communities. The DMS is designed as a nonlinear, relational, and educator-driven framework, emphasising that decolonisation is a continuous, iterative practice rooted in collaboration and self-reflection.

\subsection{The Seven Layers}
Table \ref{tab:dms_layers} summarises the seven layers of the DMS, detailing their stages, descriptions, and relational lenses. These layers provide a structured yet flexible approach to integrating Indigenous perspectives into CE.

\begin{table*}[htbp!]
\centering
\small
\caption{The Decolonial Mindset Stack (DMS): Mindset Layers and Relational Lenses in Computing Education}
\begin{tabular}{p{4.7cm} p{2cm} p{7.5cm} p{2.1cm}}
\toprule
\textbf{Mindset Layer} & \textbf{Stage} & \textbf{Description} & \textbf{Relational Lens} \\
\midrule
\textbf{Layer 7: Sovereignty Layer} & Resurgence & Indigenous communities lead and govern digital infrastructure, data, and algorithmic systems for their own futures. & \textbf{By Us} \\
\addlinespace
\textbf{Layer 6: Knowledge Recovery Layer} & Reclamation & Respectful emulation of Indigenous ways of knowing, being, and doing, guided by Elders and Indigenous scholars. & \textbf{Like Us} \\
\addlinespace
\textbf{Layer 5: Relational Practice Layer} & Reciprocity & Ongoing relational, reciprocal, and ethical learning partnerships with Indigenous knowledge holders and communities. & \textbf{Between Us} \\
\addlinespace
\textbf{Layer 4: Embedding Layer} & Reembedding & Co-creation and intentional embedding of Indigenous content, methods, and protocols in curriculum and teaching. & \textbf{For Us} \\
\addlinespace
\textbf{Layer 3: Conceptual Redesign Layer} & Reframing & Dislodging colonial logic in CE and preparing a space for ethical inclusion of Indigenous perspectives. & \textbf{With Us} \\
\addlinespace
\textbf{Layer 2: Reflection Layer} & Reflection & Personal and institutional examination of internalised colonial assumptions and complicity in exclusionary structures. & \textbf{Of Us} \\
\addlinespace
\textbf{Layer 1: Awareness Layer} & Recognition & Developing awareness of First Peoples’ histories, cultures, sovereignty, and their ongoing exclusion from computing systems. & \textbf{About Me} \\
\bottomrule
\end{tabular}
\vspace{0.5em}
\begin{quote}
\footnotesize
\textbf{Note:} Each mindset layer is relationally grounded and non-linear. This stack scaffolds ethical engagement and Indigenous-led transformation in CE.
\end{quote}
\label{tab:dms_layers}
\end{table*}


\textbf{Layer 1: Recognition (\textit{About Me})}: 
This stage focuses on acknowledging and uncovering the historical and current impacts of colonisation in CE. Through the ``About Me'' lens, learning about Indigenous contributions and perspectives sets the foundation. Recognition initiates the decolonial process by fostering awareness of First Peoples’ histories, cultures, and ongoing contributions to computing and technology. This layer encourages educators to acknowledge the historical exclusion of Indigenous voices and to lay the groundwork for inclusive education. Activities include conducting curriculum audits to assess the presence of Indigenous perspectives in course materials and hosting storytelling sessions where Indigenous knowledge holders share their experiences and insights. Outcomes might include curriculum audits, awareness raising, and explicit recognition of Indigenous knowledge as valuable.

\textbf{Layer 2: Reflection (\textit{Of Us})}: 
The ``Of Us'' lens invites seeing common humanity while respecting difference; recognising that mainstream teaching often concentrates on non-Indigenous norms.
Reflection involves critical self-examination of personal biases, privileges, and complicity in perpetuating colonial structures within educational settings. Educators are prompted to evaluate how their teaching practices may reinforce exclusion and to consider pathways for transformation. Activities include guided journaling to explore individual assumptions and facilitated Yarning Circles, which provide a culturally safe space for collective dialogue and reflection. Expected outcomes include a personal shift towards valuing multiple worldviews and preparing to integrate new approaches.

\textbf{Layer 3: Reframing (\textit{With Us})}: 
Reframing centres on dismantling colonial frameworks embedded in both curriculum content and pedagogical approaches, making space for Indigenous perspectives. This layer challenges educators to rethink and redesign how computing is taught to align with decolonial values. Activities include syllabus surgery, where course content is critically reviewed and revised, and bias-spotting code reviews, where students and instructors examine code for cultural biases or exclusionary assumptions. Outcomes may include consultations with First Nations Elders, professors, and professionals, increased student engagement, development of culturally sensitive technologies, and a more nuanced understanding of the role of computing in society. 

\textbf{Layer 4: Reembedding (\textit{For Us})}: 
Reembedding emphasises the co-creation of curriculum with Indigenous partners, ensuring that educational content and methods are culturally relevant and beneficial to Indigenous communities. This layer integrates Indigenous knowledge systems into CE collaboratively. Activities include design jams, where educators and Indigenous contributors develop new course modules together, and on-country learning modules that connect computing concepts to Indigenous land stewardship practices. Refembedding may involve new course modules on Indigenous technologies, project-based learning with community impact, and inclusive teaching practices. Success might include new programs or courses explicitly targeting Indigenous contexts and measurable increases in Indigenous student participation.

\textbf{Layer 5: Reciprocity (\textit{Between Us})}: 
Reciprocity engages in establishing ongoing, equitable partnerships with Indigenous communities, built on mutual respect and shared benefits. This layer underscores the importance of sustained relational accountability in educational initiatives. Activities include formalising partnerships through Memorandums of Understanding (MOUs), engaging in co-mentored research projects, and implementing community feedback loops to align teaching with community priorities. Indicators of success include long-term partnerships, co-developed research, and evidence of knowledge flow in both directions.

\textbf{Layer 6: Reclamation (\textit{Like Us})}: 
Reclamation involves the respectful emulation of Indigenous ways of knowing, being, and doing, led by Indigenous mentors and scholars. Educators adopt Indigenous pedagogies and practices to enrich CE ethically. Activities include mentor-led knowledge workshops, where Indigenous Elders share cultural teachings, and the integration of cultural protocols into computing projects to ensure respectful engagement. Examples include incorporating traditional knowledge into computing projects, using Indigenous languages or examples, and supporting students to reclaim their heritage. Validation could be seen in curriculum materials being co-developed with Indigenous partners and positive feedback from communities.

\textbf{Layer 7: Resurgence (\textit{By Us})}: 
Resurgence envisions Indigenous self-determination, where Indigenous communities lead and govern their own digital infrastructure, data, and algorithmic systems. This layer empowers Indigenous leadership in shaping the future of CE and technology. Activities include supporting the development of community-owned platforms and crafting data sovereignty policies that prioritise Indigenous control over their digital resources.  Curriculum and community efforts spark a broader cultural revitalisation, inspiring more Indigenous students to pursue computing. Illustrative validation is an increase in Indigenous-led tech initiatives, leadership roles, and a computing culture that fully integrates Indigenous worldviews.

\subsection{Positional Lens and Educator-driven Process}
The relational lenses in the DMS show educators in shifting their perspectives and relationships with Indigenous communities across each layer. In the Awareness Layer, ``About Me'' prompts educators to develop self-awareness regarding their own positionality and how they relate to First Peoples’ histories and ongoing exclusion. ``Of Us'' in the Reflection Layer encourages a collective examination of internalised colonial assumptions within educational institutions. ``With Us'' in the Conceptual Redesign Layer emphasises collaboration with Indigenous communities to dismantle colonial frameworks in CE. ``For Us'' in the Embedding Layer aims at co-creating curricula that serve the needs and aspirations of Indigenous communities. ``Between Us'' in the Relational Practice Layer highlights the establishment of ongoing, reciprocal partnerships. ``Like Us'' in the Knowledge Recovery Layer involves learning and adopting Indigenous ways of knowing, being, and doing under the guidance of Elders and Indigenous scholars. Finally, ``By Us'' in the Sovereignty Layer supports Indigenous communities in leading and governing their own digital infrastructure, data, and algorithmic systems, ensuring self-determination and sovereignty.

The DMS is nonlinear, allowing educators to revisit and iterate through layers as they deepen their understanding and relationships with Indigenous communities. This relational process hinges on ongoing dialogue and collaboration with Indigenous knowledge holders, ensuring that decolonisation remains a dynamic, reciprocal practice. As an educator-driven approach, the DMS places the responsibility on educators to initiate and sustain this transformation through their own learning, reflection, and commitment to change.

These activities serve as illustrative entry points, not prescriptive mandates. Educators must adapt them to the specific cultural and historical contexts of the Indigenous communities they serve, ensuring alignment with local priorities \cite{taylor2018new}. Collaborative design with Indigenous knowledge holders, Elders, and organisations is essential to maintain cultural appropriateness and foster reciprocal relationships. To demonstrate, Australian practitioners adopted the 8-Ways and yarning circles in indigenising an IT curriculum at the University of Canberra \cite{romano2023indigenizing}, while a Canadian scholar reshaped CE with Two-Eyed Seeing at the University of Ottawa \cite{habash2024two}. Interestingly, Karetai showcased practices in decolonising CE with Māori's approach in New Zealand \cite{karetai2023decolonising}. Overall, the nonlinear nature of the DMS allows educators to revisit and refine these activities as they deepen their engagement with local Indigenous communities.

\section{Discussion}
\label{sec:discussion}
This section explores the significance, scope, audience, ethics and reflections of implementing the DMS. The institutional barriers that may hinder its adoption, practical applications in educational settings, and potential avenues for future validation. By tackling these concerns, the DMS will foster a more inclusive and equitable CE landscape that honours Indigenous knowledge and perspectives.

\begin{figure}[ht!]
    \centering
    \includegraphics[width=0.45\textwidth]{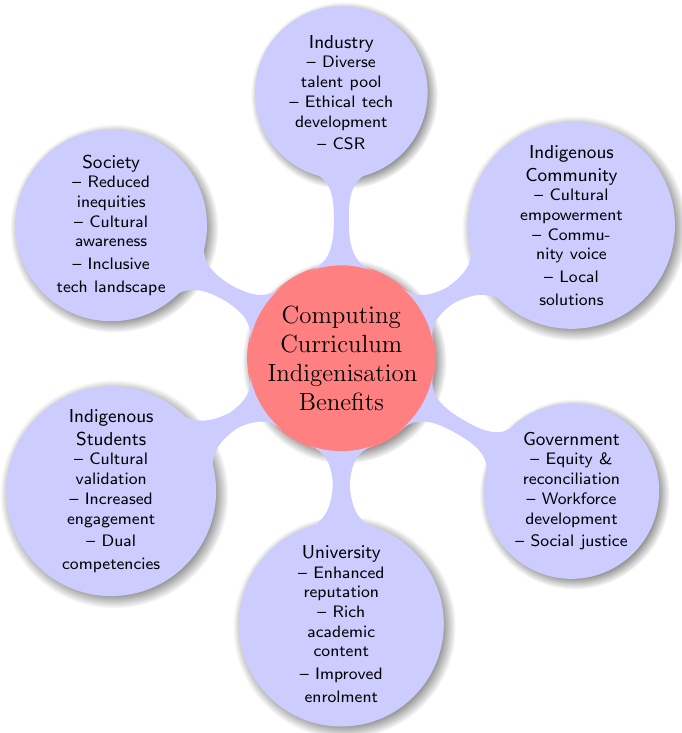}
    \caption{Computing Curriculum Indigenisation Benefits for key Stakeholders}
    \label{fig:benefit}
\end{figure}

\textbf{Significance}: 
Universal mastery of computing technologies is non-negotiable for future empowerment, but achieving true universality requires culturally inclusive education through thoughtful indigenisation. It's not just about learning how to code; it's about learning why and for whom we code, ensuring the ``future world'' powered by technology is built by and beneficial for all of humanity, reflecting its rich diversity \cite{atabey2025fairness}. This approach is fundamental to creating a more equitable, innovative, and sustainable technological future \cite{li2023current}. In this regard, Fig. \ref{fig:benefit} showcases benefits for primary stakeholders in indigenising computing curricula.

\textbf{Scope and Limitations}:
The DMS was developed with reference to First Peoples in Australia, but the issue of Indigenous underrepresentation and the colonial legacy of CE is a global concern. Admittedly, Indigenous communities are not monolithic; First Peoples have heterogeneous histories, culture, worldviews, and pedagogies \cite{sissons2005first}, and this paper only lists a few. However, local adaptation and co-leadership are essential. While the stages are sequenced conceptually, they are not strictly linear and may recur or overlap depending on context, readiness, and relational dynamics.

\textbf{Intended Audience and Relational Readiness}: 
The technical focal point of CE, often rooted in Eurocentric frameworks, prioritises abstract concepts like algorithms over Indigenous knowledge systems, hindering inclusive curriculum development \cite{armoni2014early}. The DMS sees to this challenge by offering a layered, relational framework to guide educators in transforming CE. Drawing inspiration from the ISO OSI model, which progresses from physical networks to user applications \cite{nelson2000teaching}, the DMS scaffolds educators’ mindsets from awareness of Indigenous exclusion to supporting Indigenous sovereignty.

The DMS targets non-Indigenous computing educators committed to fostering Indigenous inclusion and sovereignty. As a conceptual escort, it does not prescribe rigid steps but supports ethical, culturally responsive transformations tailored to diverse Indigenous contexts. Central to the framework is \textbf{relational readiness}---the willingness to critically examine personal biases, share power, and build reciprocal, accountable partnerships with Indigenous communities \cite{brant2023partnership}. This mindset shift is a prerequisite for meaningful indigenisation, ensuring CE respects and amplifies Indigenous voices towards improved representation.

\textbf{Institutional Constraints and Ethical Risks}: 
Decolonising CE encounters institutional barriers like inertia and tokenism, where universities endorse Indigenous inclusion through symbolic policies but underfund initiatives or misalign with community needs \cite{omodan2024roles}. For example, adding a single Indigenous module without broader curriculum reform risks superficiality. The DMS counters these by promoting sustained, relational transformation across diverse Indigenous contexts \cite{gaudry2018indigenization}. To overcome resistance to changing computing curricula and educators’ lack of training, institutions must invest in professional development, fund partnerships, and recognise decolonial efforts in evaluations, ensuring ethical, accountable indigenisation that aligns with Indigenous cultural and sovereignty goals.

\textbf{Equity, Sovereignty, and Justice}: 
Indigenising CE seeks epistemic justice, i.e., fair recognition of Indigenous knowledge systems, challenging colonial biases that prioritise Western frameworks in algorithms and curricula. The DMS pivots Indigenous sovereignty, ensuring communities shape their digital futures, such as through data governance. Ethical implementation requires relational accountability via reciprocal partnerships and adherence to principles like respect and consent, as outlined in ethical guidelines \cite{tomkins2024aiatsis}. Continuous collaboration with Indigenous knowledge holders across global contexts, from Māori to Navajo \cite{yan2024teaching}, ensures authentic integration, fostering equitable CE that amplifies Indigenous voices and drives innovation.

\textbf{Sustaining the Work}: 
The DMS supports practical applications in CE, guiding educators through workshops, curriculum redesign, and professional development. Workshops use activities like curriculum audits to assess Indigenous content and Yarning Circles, a culturally safe dialogue practice, to reflect on biases. Curriculum redesign integrates Indigenous perspectives, such as case studies on Indigenous-led AI projects or on-country coding modules connecting to environmental stewardship \cite{gaudry2018indigenization}. Professional development embeds DMS layers to foster decolonial mindsets, enhancing culturally responsive teaching. Sustaining this work requires Indigenous-led collaboration, supported by institutional structures like formal partnerships and compensated community review of curricula, adhering to ethical guidelines \cite{tomkins2024aiatsis}. The DMS’s relational lenses ensure adaptability to diverse Indigenous contexts, enabling scalable, community-driven indigenisation that fosters equitable CE globally.

\textbf{Reflection in Practice}: 
To support educators in engaging with the DMS, we suggest the following guiding questions:

\begin{itemize}
    \item Who is missing from your curriculum, and why?
    \item What assumptions about knowledge and technology are embedded in your teaching?
    \item How do you engage with Indigenous perspectives ethically and relationally?
    \item Who benefits from your curriculum as it currently stands?
\end{itemize}
These questions serve as entry points to the mindset transformation that the DMS scaffolds, inviting educators to locate themselves in the broader work of decolonising CE \cite{gaudry2018indigenization}. 

Although this paper is conceptual, future research could validate the DMS through pilot projects in CE settings. These projects might involve implementing the DMS in specific courses or departments, evaluating its impact on educator practices, student engagement, and community relationships through surveys, interviews, or curriculum analyses. Such empirical studies would provide insights into refining the framework and demonstrating its potential to foster equitable CE, paving the way for broader adoption.

\section{Conclusion}
\label{sec:conclusion}
Computing education (CE) holds transformative potential for advancing economic mobility, cultural preservation, data sovereignty, and self-determination. For Indigenous communities, participation in computing must move beyond mere access towards genuine empowerment and control. Addressing the persistent underrepresentation of First Peoples requires more than content inclusion, it demands a fundamental reorientation of educator mindsets, institutional structures, and pedagogical relationships. In response, we introduced the DMS—a conceptual, layered framework informed by relational lenses and grounded in Indigenous methodologies and Freirean critical pedagogy. The DMS scaffolds educators through seven mindset transformations: Recognition, Reflection, Reframing, Reembedding, Reciprocity, Reclamation, and Resurgence. 

Adopting a decolonial mindset is essential to transforming CE in ways that confront the historical and ongoing exclusion of Indigenous peoples. 
Central to this transformation are the educator competencies of relationality and reflexivity. Relationality involves building genuine, reciprocal relationships with Indigenous communities, ensuring their voices and knowledge systems are central to curriculum design and delivery. This approach positions education as a collaborative dialogue rather than a one-directional transfer of knowledge. Reflexivity requires educators to critically examine their own biases, privileges, and roles in maintaining colonial structures, committing themselves to ongoing learning and accountability. Together, these competencies equip educators to navigate the complexities of decolonisation with cultural sensitivity and ethical responsibility.

We call upon computing educators to engage in co-transformative work with Indigenous communities, using frameworks like the DMS to guide practice. By adopting relational and reflexive approaches, educators can co-create curricula that reflect Indigenous histories, cultures, and aspirations, fostering a more just, inclusive, and sustainable future for CE. This work addresses historical injustices as well as unlocks diverse perspectives that are vital for innovation.

While the DMS has been developed within the context of CE, we believe its principles and processes are applicable across a range of technological disciplines. The journey of decolonisation is ongoing and requires sustained commitment and collaboration, but it is essential to ensuring that all learners are empowered to shape the technological landscapes of the future.

\section*{Acknowledgement}
We acknowledge the Traditional Custodians of the lands on which this work was undertaken and pay our respects to Elders past and present. We thank the Indigenous knowledge holders and communities who have guided this work, including Prof Mark Rose, Prof Gabrielle Fletcher, A/Prof Al Fricker, A/Prof Cassandra Seery, and Ms Joleen Ryan (Deakin University), as well as Sarah Watts and Cody Blundstone (Indigital). Their contributions are central to reimagining computing education. This research was supported by the ALTA Grant, funded by ACDICT, and by the Indigenous Storytelling Project on Cybersecurity Awareness among First Peoples, funded by Deakin Cyber (PJ08575).

\bibliographystyle{elsarticle-num}
\bibliography{toce}

\end{document}